\documentclass[conference]{IEEEtranSPMB}

\IEEEoverridecommandlockouts
\ifCLASSINFOpdf
\else
\fi
\usepackage{amsmath,graphicx,soul,subcaption,balance}
\usepackage{balance}
\usepackage[numbers,sort&compress]{natbib}

\usepackage{parskip}
\usepackage{float} 

\usepackage{mathptmx}
\usepackage{hyperref}
\usepackage{caption}

\usepackage{multirow} 

\usepackage{color}
\usepackage{soul,mathrsfs}
\usepackage[utf8]{inputenc}
\usepackage[T1]{fontenc}
\usepackage{array}

\usepackage[papersize={8.5in,11in}, left=1in, right=1in, top=1in, bottom=1in]{geometry}
\newcolumntype{L}[1]{>{\raggedright\let\newline\\\arraybackslash\hspace{0pt}}m{#1}}
\newcolumntype{C}[1]{>{\centering\let\newline\\\arraybackslash\hspace{0pt}}m{#1}}
\newcolumntype{R}[1]{>{\raggedleft\let\newline\\\arraybackslash\hspace{0pt}}m{#1}}

\usepackage[singlelinecheck=false,justification=centering]{caption}
\DeclareCaptionLabelFormat{mylabel}{#1 #2.\hspace{0.7ex}}
\renewcommand{\thetable}{\arabic{table}}

\usepackage{xcolor,colortbl}
\usepackage{graphicx}
\definecolor{light-gray}{gray}{0.83}

\newcommand{\spmbtitlefont}{\large\bf\vspace{2em}}

\newcommand{\subparagraph}{}
\usepackage[compact]{titlesec}
\titleformat{\section}
   {\center\normalfont\sc}{\thesection.}{0.7ex}{}
\titlespacing{\section}{0pt}{2ex}{1.5ex}
\titlespacing{\subsection}{0pt}{1.5ex}{1.2ex}
\titlespacing{\subsubsection}{0pt}{1ex}{0.9ex}
\makeatletter
\renewcommand*{\@seccntformat}[1]{\csname the#1\endcsname .\hspace{0.7em}}
\makeatother

\usepackage{fancyhdr,lastpage}
\fancypagestyle{firststyle}
{
   \fancyhf{}
	\lfoot{IEEE SPMB 2020}\rfoot{v2.0: Oct 15th, 2020}
}
\usepackage{graphicx}
\usepackage{tikz}
\tikzstyle{process} = [rectangle, minimum width=3cm, minimum height=1cm, text centered, text width = 6cm, draw=black, fill=orange!30]
\tikzstyle{arrow} = [thick,->,>=stealth]

\title{\spmbtitlefont Machine Learning Applications to DNA Subsequence and Restriction Site Analysis

{\vspace{-2.3\baselineskip}
}
}

    \author{\IEEEauthorblockN{
    E. Moyer\textsuperscript{\it 1} and
    A. Das\textsuperscript{\it 2}
    }
    \vspace{0.5em}
    \IEEEauthorblockA{
        1. School of Biomedical Engineering, Science and Health Systems, Drexel University, Philadelphia, Pennsylvania, USA \\
        2. College of Engineering, Drexel University, Philadelphia, Pennsylvania, USA \\
        \{ejm374, anup.das\}@drexel.edu
    }
}

\newcommand{\AbstractSummary}{E.\ Moyer et al.: Machine learning ...}

\begin{document}

\IEEEaftertitletext{}
\maketitle

\begin{abstract}
Based on the BioBricks™ standard, restriction synthesis is a novel catabolic iterative DNA synthesis method that utilizes endonucleases to synthesize a query sequence from a reference sequence. In this work, the reference sequence is built from shorter subsequences by classifying them as applicable or inapplicable for the synthesis method using three different machine learning methods: Support Vector Machine (SVM), random forest, and Convolution Neural Network (CNN). Before applying these methods to the data, a series of feature selection, curation, and reduction steps are applied to create an accurate and representative feature space. Following these preprocessing steps, three different pipelines are proposed to classify subsequences based on their nucleotide sequence and other relevant features corresponding to the restriction sites of over 200 endonucleases. The sensitivity using SVM, random forest, and CNN are 94.9\%, 92.7\%, 91.4\%, respectively. Moreover, each method scores lower in specificity with SVM, random forest, and CNN resulting in 77.4\%, 85.7\%, and 82.4\%,  respectively. In addition to analyzing these results, the misclassifications in SVM and CNN are investigated. Across these two models, different features with a derived nucleotide specificity visually contribute more to classification compared to other features. This observation is an important factor when considering new nucleotide sensitivity features for future studies.
\end{abstract}

{\textbf{\textit{keywords: machine learning, subsequence matching, DNA sequence classification}}} 
\IEEEpeerreviewmaketitle    
\thispagestyle{firststyle}  

\section{Introduction}
\label{sec:intro}
Originally postulated by Watson and Crick in the 1950s, the process of cellular deoxyribonucleic acid (DNA) synthesis has long been understood as a trademark to biological studies \cite{ref1} \cite{reichard1988interactions}. Over the past few decades, understanding this process has led to the discovery and development of many synthetic manufacturing methods, namely synthetic gene synthesis \cite{ref2} \cite{caruthers1985gene}. As a result, other areas of genomics have accumulated massive research interest. From the start of the Human Genome Project in 1990 to the discovery of the Cluster Regularly Interspaced Palindromic Repeat (CRISPR) Cas9 system in bacteria during 2005, the field of genomics has advanced at an incredible rate as the reservoir of information pertaining to DNA and gene research increases~\cite{ref3}~\cite{ref4}. Although the public adoption of these technologies is in the near future, the cost of producing novel genes remains as the field’s largest determinant~\cite{ref5}.

\begin{figure}[H]
    \centering
    \includegraphics[width=0.8\linewidth]{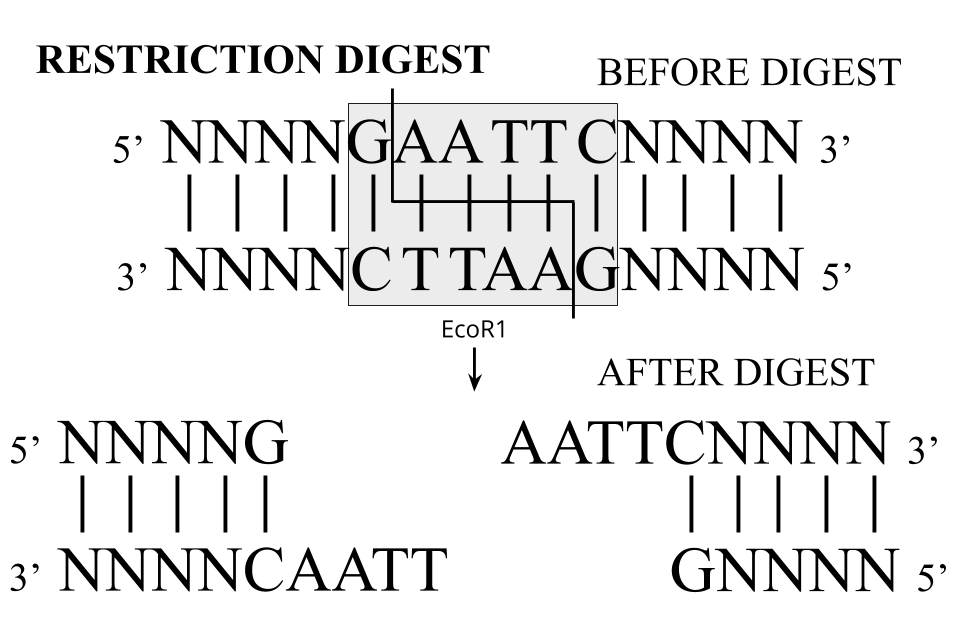}
    \setlength{\belowcaptionskip}{-15pt}
    \caption{Restriction enzyme digest of GAATTC using EcorR1. In general, DNA is represented as a permutation of the four nucleotides [adenine (A), thymine (T), cytosine (C), guanine (G)] and an ambiguous base (N).}
    \label{dnadigest}
\end{figure}

Stemming from the need to decrease the cost of synthetic synthesis, the rationale behind this research begins with the introduction of a hypothesized catabolic synthesis process, restriction synthesis. This procedure relies on a process known as restriction enzyme digest. This typical laboratory procedure utilizes restriction endonucleases, which are enzymes that cleave DNA at specific restriction sites, or target locations. \cite{bigger1973recognition}. \autoref{dnadigest} shows an example restriction enzyme digest using enzyme EcoR1 on restriction site GAATTC.

\begin{figure}[H]
    \centering
    \includegraphics[width=0.8\linewidth]{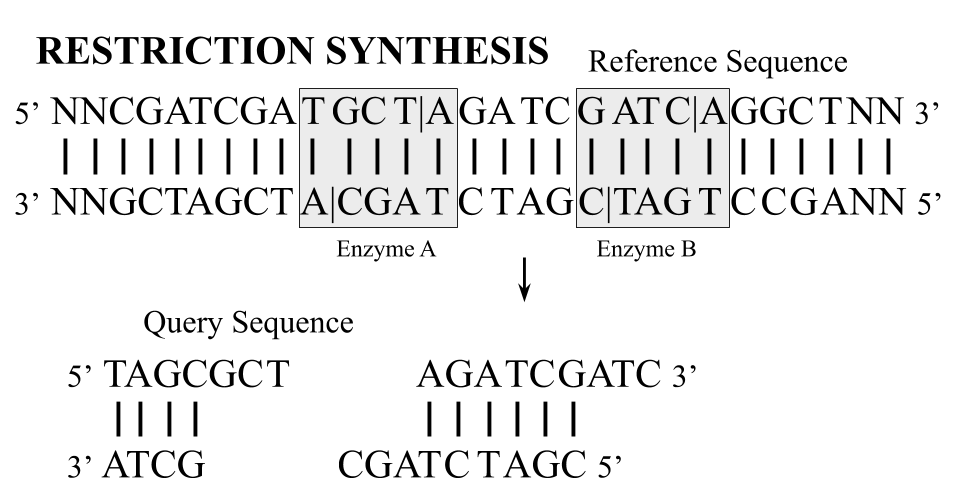}
    \setlength{\belowcaptionskip}{-15pt}
    \caption{Restriction Synthesis. Reference sequence is flanked by Enzyme A and Enzyme B. This fragment is then ligated to the end of the query sequence.}
    \label{restrictionsyn}
\end{figure}

Although restriction enzyme digest is normally reserved for a few endonucleases at a time, restriction synthesis serializes the use of many endonucleases to digest a reference sequence into fragments. These fragments are then used to synthesize a query sequence by ligating them to one another in succession as shown in \autoref{restrictionsyn}. This process is a modification to the BioBricks™ standard proposed by Dr. Thomas Knight. Anderson et al. defines the BioBricks™ standard as a process that “employs iterative restriction enzyme digestion and ligation reactions to assemble small basic parts into larger composite parts” \cite{ref7}. These larger components are built using fragments flanked by two restriction enzymes, Xbal and SpeI. The key difference between restriction synthesis and the BioBricks™ standard is that the latter only digests fragments with XbaI and SpeI endonucleases. Conversely, restriction synthesis relies on a data set of approximately 200 endonucleases from New England Biolabs \cite{ref8}. In flanking a reference sequence with many endonucleases, restriction synthesis builds more composite fragments homologous to those referenced by the BioBricks™ standard~\cite{shetty2008engineering}.

One assumption of restriction synthesis is that the reference sequence and the query sequence are both known prior to synthesis. If each query sequence is synthesized from an independent reference sequence, this assumption would require costly sequencing. Although this one-to-one correspondence between the query sequence and the reference sequence would result in incredibly accurate synthesis, it would massively increase the cost of this method. A solution is to build one reference sequence from which many different query sequences can be synthesized. While this solution is a less expensive alternative to the former, it requires implementing a procedure to carefully choose which subsequences should be included in the reference sequence. An applicable subsequence is defined as containing a candidate fragment that can be flanked by two endonucleases. It follows that subsequences can be classified based on whether they are applicable (true) or inapplicable (false) for inclusion in the reference sequence.

The data set for this research is obtained through the synthesis of mycobacterium structural genes during restriction synthesis. Each gene sequence ranges from 1,000 to 5,000 base pairs and is obtained from the National Center for Biotechnology Information (NCBI)~\cite{NCBI2020nuc}. Using this data, the analysis of subsequences is explored using three popular techniques in machine learning and neural network architecture: Support Vector Machine (SVM), random forest, and Convolution Neural Network (CNN) \cite{ref9}.

\section{Feature Selection}

The feature set for each subsequence entry includes 16 unique variables. The ‘SEQ’ feature is responsible for capturing the nucleotide sequence of each entry formatted based on the following convention: nucleotides A, T, C, and G are assigned ordinal values 1/4, 1/2. 3/4, and 1, respectively, and the ambiguous base, N, is assigned an ordinal value of 0. This feature most directly relates whether a specific sequence can be flanked by two enzymes due to the restriction sites of each endonuclease. Another feature, ‘LEN,’ describes the length of each sequence, which is related to the dimension of the 'SEQ' feature.

Eight other features are obtained using \autoref{propnuc}, which describes the proportion of each nucleotide in a sequence. Four of the eight features describes the proportion of each nucleotide in the subsequence, and the other set of four describes that of the reference sequence. 

The next set of four features are related to a novel sequence metric, the complexity rating, as shown by \autoref{comprating} and \autoref{proprating}. These equations are used to accurately describe the nucleotide composition with a continuous value ranging from 0 exclusive to 1 inclusive. \autoref{comprating} is responsible for determining the deviation of an entire sequence from an equal nucleotide composition, whereas \autoref{proprating} segments a sequence based on preset values ranging from $b$ to $p$ and examines this deviation for each segment proportionally. In other words, the rational behind \autoref{proprating} is to sum complexity ratings from \autoref{comprating} in proportion to the number of occurrences of segment size $i$ in sequence size $n$. In short, both of these ratings determine how far a sequence deviates from an equal nucleotide composition. The four features represent both the $r_1$ and $r_2$ values of the subsequence and reference sequence.

\begin{figure}[H]
    \begin{equation}\label{propnuc}
        p(x, L)=\frac{occurrence\ of\  x}{L}
    \end{equation}
    \begin{equation}\label{comprating}
    \begin{split}
       r_1 (L)= 1 - \sum_{i\in\alpha}{(\frac{1}{4} - p(i, L)})^2
    \end{split}
    \end{equation}
     \begin{equation}\label{proprating}
       r_2 (n) = \sum_{i = b}^{p} \frac{\sum_{j = 0}^{n - i} r_1(i) * \frac{i}{n}}{\sum_{k = b}^{p} \frac{k}{n}}
    \end{equation}
    \caption{Relevant equations for features describing the nucleotide proportion and complexity ratings of both the subsequence and the query sequence. Note: $\alpha$ = {\{A,T,C,G\}}}
    \label{figconvequ}
\end{figure}
\vspace{-1em}
Two additional auxiliary features, ‘C’ and ‘EZY,’ are included to describe the number of query sequences in a specific list that contain the subsequence and the number of available enzymes that can digest the subsequence, respectively. These are incorporated to gauge whether a particular subsequence is common in the data set.

\section{Feature Analysis}
Feature reduction is performed in order to simplify the feature space for each of the three machine learning techniques. A correlation test is applied across all features to determine whether any of them are redundant. A conservative threshold is set at 0.90 because of the small size of the feature set. None of the pairs of features had a correlation higher than 0.90, so all features are included in the study. A sample feature correlation test on the first six features can be observed in \autoref{corrtest} below.

\begin{table}[H]
    \centering
    \includegraphics[width=0.9\linewidth]{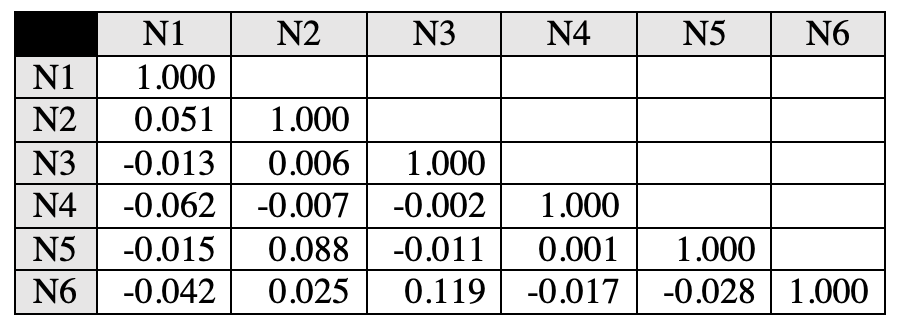}
    \setlength{\belowcaptionskip}{-15pt}
    \caption{An example correlation test on the first six nucleotides (N1 to N6) for the 'SEQ' feature.}
    \label{corrtest}
\end{table}

Surprisingly, this indicates low nucleotide-nucleotide interaction within a subsequence. One would expect some degree of interaction due to the nature of restriction enzymes cutting at specific restriction sites.

\section{Data Selection \& Curation}

A constant training set and testing set is used while evaluating each of the machine learning solutions. Based on an initial raw data set of approximately one million entries across almost one hundred gene synthesis simulations, a sample of 105,000 entries is selected with a 50/50 percent distribution of applicable and inapplicable subsequences. From that data set, a stratified sample produces both the training set and testing set with unique entries. The training set consists of 50,000 unique entries with a 60/40 percent distribution of the classes, and the testing set consists of 26,622 unique entries with a 50/50 percent distribution of the classes.

\section{Metrics}

In this work, we compare machine learning techniques using the metrics of \autoref{sens}-7.

\begin{figure}[H]
\small
    \begin{equation}\label{sens}
        sensitivity = \frac{true\ positive}{true\ positive + false\ negative}
    \end{equation}
    \begin{equation}\label{spec}
       specificity = \frac{true\ negative}{true\ negative + false\ positive}
    \end{equation}
    \begin{equation}\label{falseneg}
       false\ negative\ rate\ (FNR) = 1 - sensitivity
    \end{equation}
    \begin{equation}\label{falsepos}
       false\ positive\ rate\ (FPR) = 1 - specificity
    \end{equation}
    \caption{Sensitivity, specificity, false negative rate, and false postive rate equations.}
    \label{figconvequ}
\end{figure}

\section{Methods}

1. Support Vector Machines (SVM)

SVM is a supervised learning model that creates hyperplanes between groups of data \cite{ref23}. Given a set of training data and corresponding labels, the algorithm attempts to best categorize data by measuring the distance of each data point to the segregating hyperplane \cite{ref24}. By maximizing these distances, SVM creates highly reliable non-probabilistic classifications \cite{ref25}.

In addition to the feature analysis, Principle Component Analysis (PCA) is also used for feature reduction. PCA is an unsupervised learning method that reduces a feature space while retaining relationships in the data. By transforming a high dimensional feature space into a specified number of principle components (PCs), raw data can be summarized in a few dimensions~\cite{ref11}.

\begin{figure}[H]
    \centering
    \includegraphics[width=\linewidth]{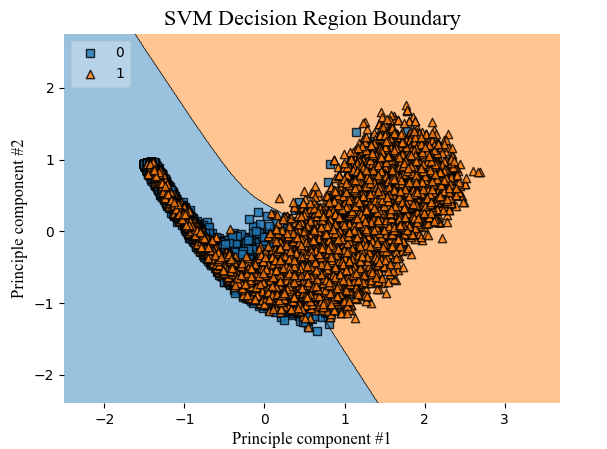}
    \setlength{\belowcaptionskip}{-15pt}
    \setlength{\abovecaptionskip}{-5pt}
    \caption{Polymetric 2 PCs SVM analysis.}
    \label{pcasvmfig}
\end{figure}

As seen in \autoref{pcasvmfig}, the classification can be easily visualized when 2 PCs are used. In this case, the most successful classifier is the 2 PCs polymetric kernel. The results from the testing set of 26,622 entries can be observed in the confusion matrix shown in \autoref{pcasvmcon}.
\vspace{-6.5px}
\vspace{-1px}
\begin{table}[H]
    \centering
    \includegraphics[width=0.8\linewidth]{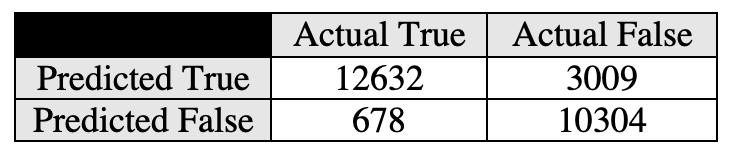}
    \vspace{-1px}
    \caption{Confusion matrix for polymetric 2 PCs SVM.}
    \label{pcasvmcon}
\end{table}
\vspace{-6px}
\autoref{pcasvmcon} displays that the sensitivity is 94.9\% and the specificity is 77.4\%. These results are seen consistently across all trials of this configuration. For this reason, the standard deviation is considered negligible. Furthermore, \autoref{pcasvmcon} indicates a 5.1\% FNR and a 22.6\% FPR.

In addition to the polymetric 2 PCs classifier, five other tests are conducted: non-PCA linear kernel, 3 PCs linear kernel, 3 PCs polymetric kernel, 2 PCs RBF kernel, and 2 PCs sigmoid kernel. The results are below in \autoref{pcaresults}.

\begin{table}[H]
    \centering
    \includegraphics[width=1.0\linewidth]{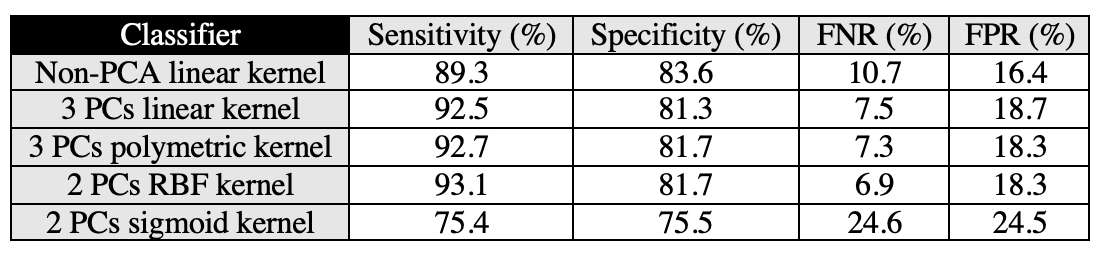}
    \setlength{\belowcaptionskip}{-15pt}
    \caption{Alternative PCA results.}
    \label{pcaresults}
\end{table}

\begin{figure}[H]
\vspace{-1em}
	\centering
	\begin{subfigure}[t]{0.8\linewidth}
		\centering
		\includegraphics[width=\linewidth]{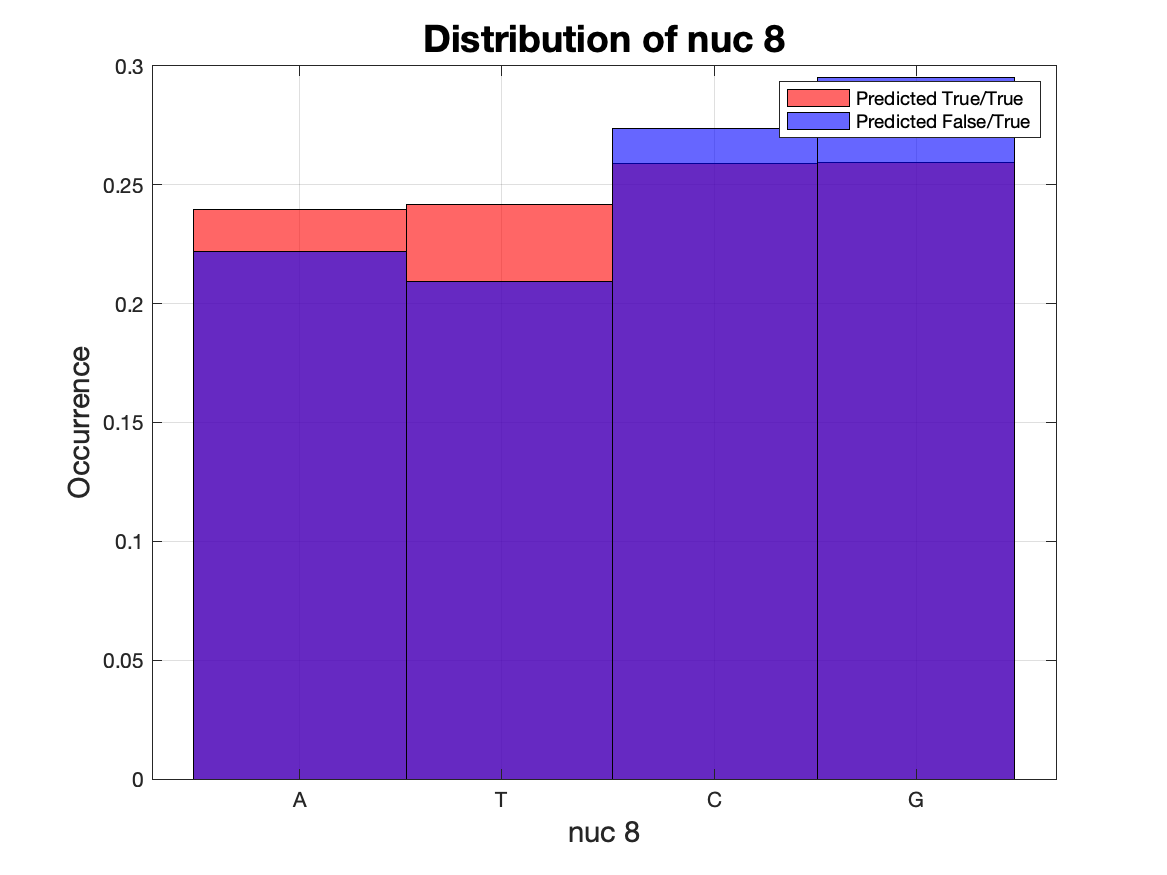}
		\caption{Distribution of the eighth nucleotide position.}\label{svmfeatures8}
	\end{subfigure}

	\begin{subfigure}[t]{0.8\linewidth}
		\centering
		\includegraphics[width=\linewidth]{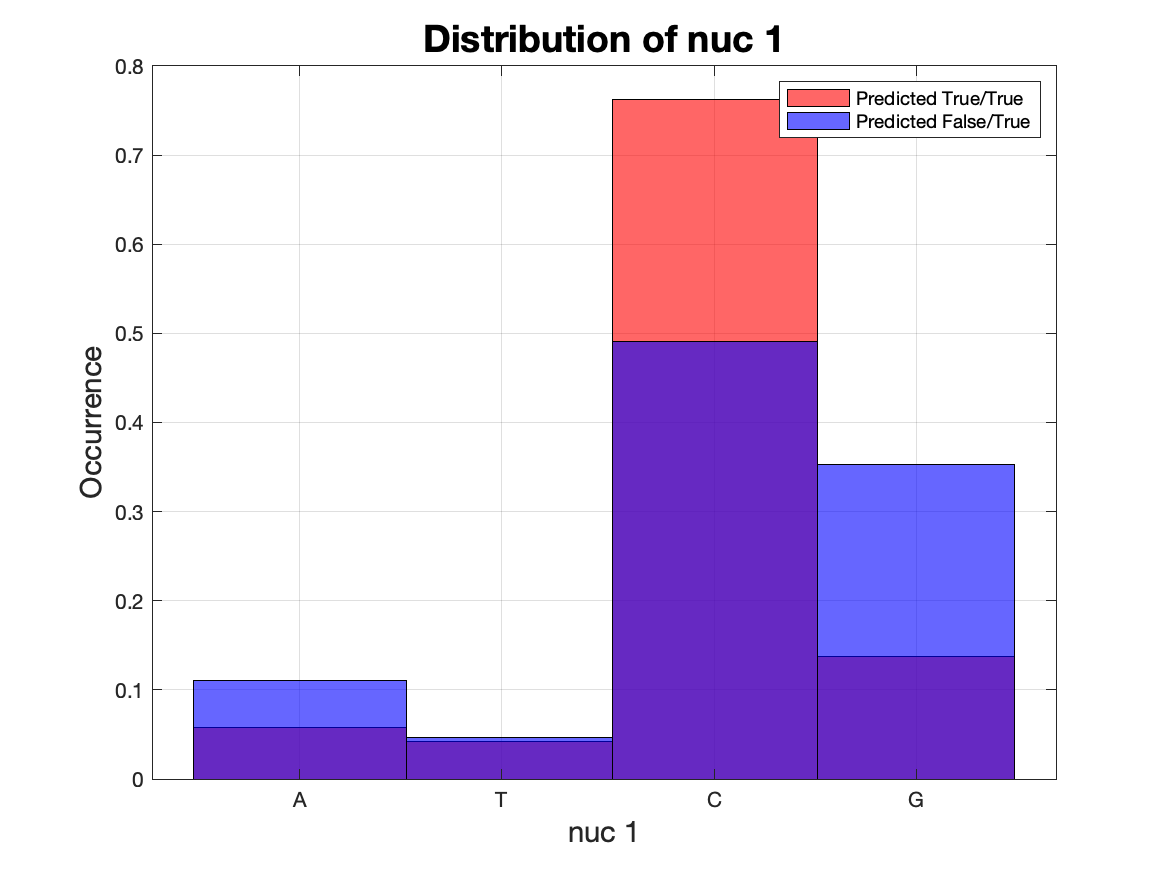}
		\caption{Distribution of the first nucleotide position.}\label{svmfeatures1}
	\end{subfigure}
	\setlength{\belowcaptionskip}{-20pt}
	\caption{Feature distribution for SVMs}\label{svmfeatures}
\end{figure}

The two histograms in \autoref{svmfeatures} display the relative proportions of each nucleotide at different features under the 'SEQ' feature. Red bars indicate correctly classified true entries, and blue bars indicate incorrectly classified true entries. Purple bars are simply the overlap of the two. Many of the features in this data set had distributions similar to the two histograms in \autoref{svmfeatures8} and \autoref{svmfeatures1}. When the distribution is seemingly uniform across all four nucleotides, such as in \autoref{svmfeatures8}, there is a displayed indifference between the four nucleotides at a specific nucleotide position. This suggests that at least across this data set, the eighth nucleotide position is insensitive to the type of base. Conversely, \autoref{svmfeatures1} displays higher base sensitivity as it favors cytosine far more than any of the other three nucleotides on the first nucleotide position. This distribution suggests the first position is incredibly sensitive to the type of base.

\vspace{10pt}

Further analysis is needed in order to determine whether these differences are statistically significant. If this is the case, it would suggest a difference in nucleotide sensitivity across nucleotide positions. 

Another interesting question arises on the misclassifications between these two features. For instance, \autoref{svmfeatures8} shows both over and under classification of true and false subsequences for the eighth nucleotide position. Further statistical analysis may suggest that there is not a significant difference between misclassification rates for each nucleotide. Additionally, \autoref{svmfeatures1} displays that the first nucleotide position misclassifies C and G in much higher proportions than A and T. Further statistical analysis may suggest this to be the case. More interestingly, these misclassifications seem to follow the distributions of correctly matched sequences. For example, \autoref{svmfeatures8} misclassified sequences uniformly across that feature. This phenomenon is also apparent in \autoref{svmfeatures1} as the proportion of misclassifications follow the distribution of true classifications for that feature.

2. Random Forest

Random forest is an ensemble learning model that implements a set number of decision trees based on subsets of training data for classification and regression~\cite{ref22}. One of the features of random forest is its use of out-of-bag data \cite{ref21}. At the construction of each tree, two-thirds of the data is included for training and the remaining one-third is set aside for validation after the training has completed \cite{ref12}. During the training process, each decision tree is generated based on a segmented data set and a randomly selected feature set at each node of the tree \cite{ref13}. While a single decision tree may overfit data, many decision trees in random forest allow for the overfitting to average out between trees. When predicting based on new data, each tree provides a prediction, or a vote, for a certain class. The final classification for random forest is based on which classification has the most votes \cite{ref22}.

The random forest analysis is conducted with 10, 20, 30, 40, and 50 trees. The most successful analysis is conducted when there were 30 trees, but even as this number passed 30, the additional level of accuracy in both categorizations began to level off. In other words, the classification accuracy when the number of trees is 50 is the same or less than when there were just 40 and 30 trees. The results from the test set of 26,622 entries can be observed in the confusion matrix in \autoref{randomforest}.

\begin{table}[H]
    \centering
    \includegraphics[width=0.8\linewidth]{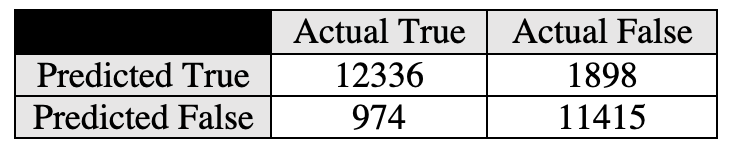}
    \setlength{\belowcaptionskip}{-15pt}
    \caption{Confusion matrix for random forest (n=30).}
    \label{randomforest}
\end{table}

With 30 decision trees, the sensitivity is 92.7\% and the specificity is 85.7\% for the testing set of 26,622 entries. These results are seen consistently across all trials of random forest. For this reason, the standard deviation is considered negligible. In \autoref{randomforest}, the confusion matrix displays a 7.3\% FNR and 14.2\% FPR.

3. Convolution Neural Networks (CNN)

CNN is a class of deep learning models that use a linear operation called convolution in at least one of their layers \cite{ref15}. The model usually begins with a convolution layer that accepts a tensor as an input with dimensions based on the size of the data \cite{ref20}. The second layer, or the first hidden layer, is formed by applying a kernel or filter that is a smaller matrix of weights over a receptive field, which is a small subspace of the inputs \cite{ref19}. Kernels apply an inner product on the receptive field, effectively compressing the size of the input space \cite{ref18}. As the kernel strides across the input space, the first hidden layer is computed based on the weights of the filter. As a result, the first hidden layer is a feature map formed from the kernel applied on the input space \cite{ref20}. While the dimension of the kernel may be much smaller in size compared to the initial inputs of the convolution layer, the kernel must have the same depth of the input space. The inputs and convolution layer are often followed by rounds of activations, normalizations, and pooling \cite{ref18}. The last layer, however, is a fully connected layer where the final outputs or categorizations are determined based on how different features fall in line with the specific classes under study \cite{ref19}.

\begin{figure}[H]
    \centering
    \begin{tikzpicture}[node distance=1.3cm]
    
    \node (rect1) [process] {Conv1D (Filters=64, Kernel size=3, Strides=3, Padding=Same, Activation=Softplus, Kernel regularizer=L2(0.01), Activity regularizer=L1(0.01))};
    
    \node (rect2) [process, below of=rect1, yshift=-0.8cm] {Conv1D (Filters=64, Kernel size=3, Strides=3, Padding=same, Activation=Linear)};

    \node (rect3) [process, below of=rect2, yshift=-0.2cm] {Dropout (Rate = 0.5)};

    \node (rect4) [process, below of=rect3]{MaxPooling1D (Pool size=2, Padding=Same)};

    \node (rect5) [process, below of=rect4] {Flatten};
    
    \node (rect6) [process, below of=rect5] {Dense (Units=128, Activation=Linear)};
    
    \node (rect7) [process, below of=rect6] {Dense (Units=2, Activation=Softmax)};
    
    \draw [arrow] (rect1) -- (rect2);
    \draw [arrow] (rect2) -- (rect3);
    \draw [arrow] (rect3) -- (rect4);
    \draw [arrow] (rect4) -- (rect5);
    \draw [arrow] (rect5) -- (rect6);
    \draw [arrow] (rect6) -- (rect7);
    
    \end{tikzpicture}
    \caption{Currently optimal CNN}
    \label{cnnmodel}
\end{figure}
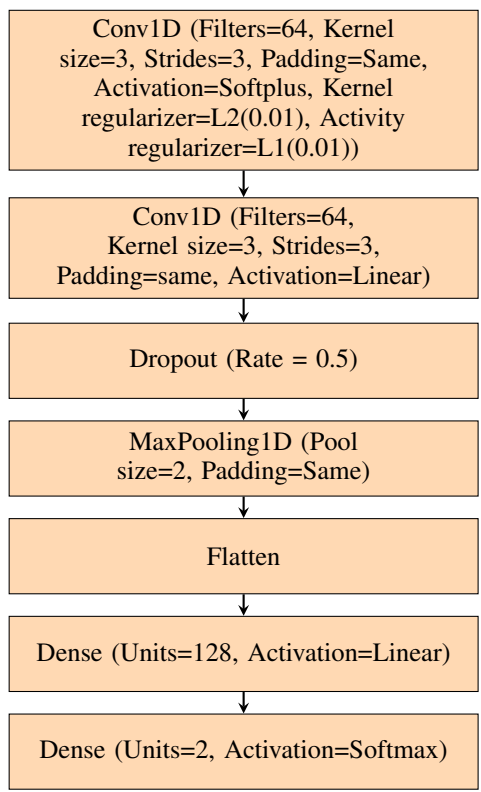

\autoref{cnnmodel} models the proposed Convolution Neural Network (CNN) for this data set.

After 100 trials of classification using this model, the average sensitivity is 91.4\% and the FNR is 8.6\% with a standard deviation of 3.5\%, and the specificity is 82.4\% and the FPR is 17.6\% with a standard deviation of 1.2\% for the testing set of 26,622 entries. 

\begin{figure}[H]
	\centering
	\begin{subfigure}[H]{0.8\linewidth}
		\centering
		\includegraphics[width=\linewidth]{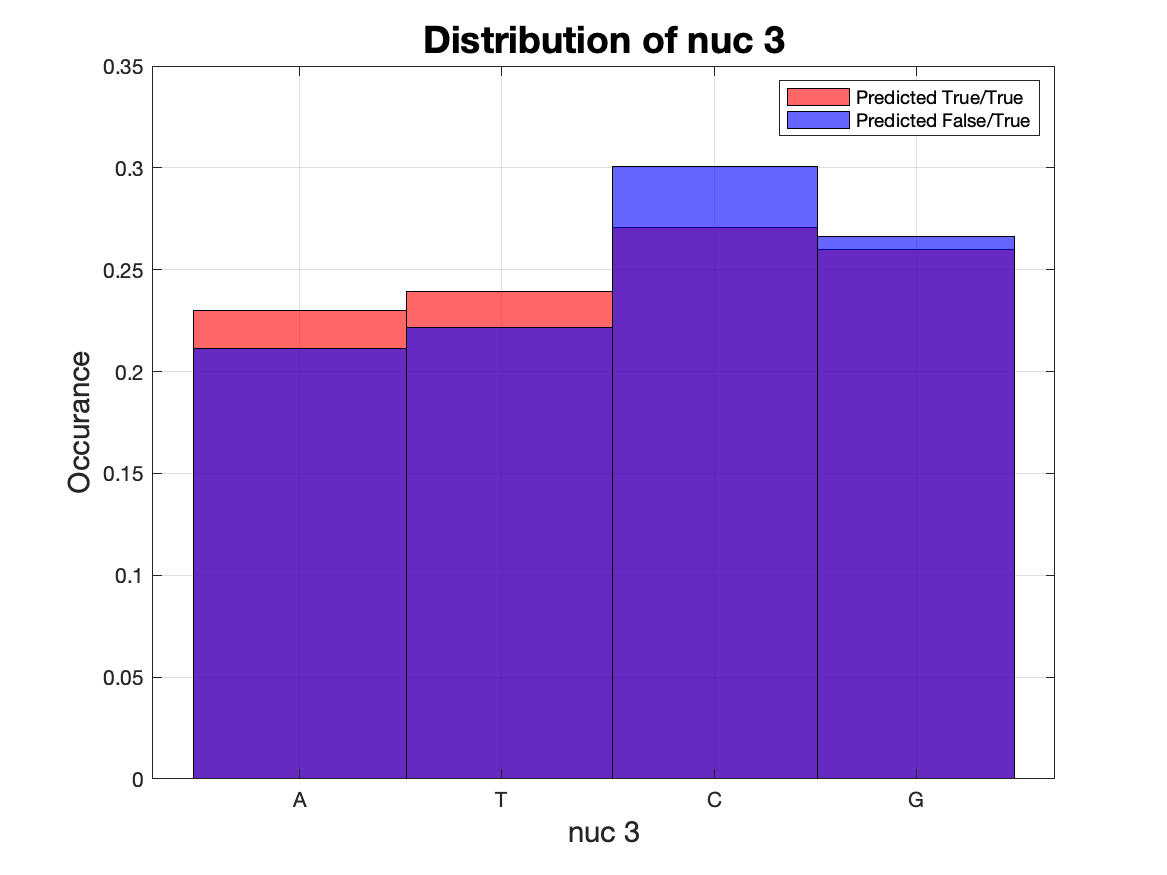}
		\caption{Distribution of the third nucleotide position.}\label{cnnfeatures3}
	\end{subfigure}

	\begin{subfigure}[H]{0.8\linewidth}
		\centering
		\includegraphics[width=\linewidth]{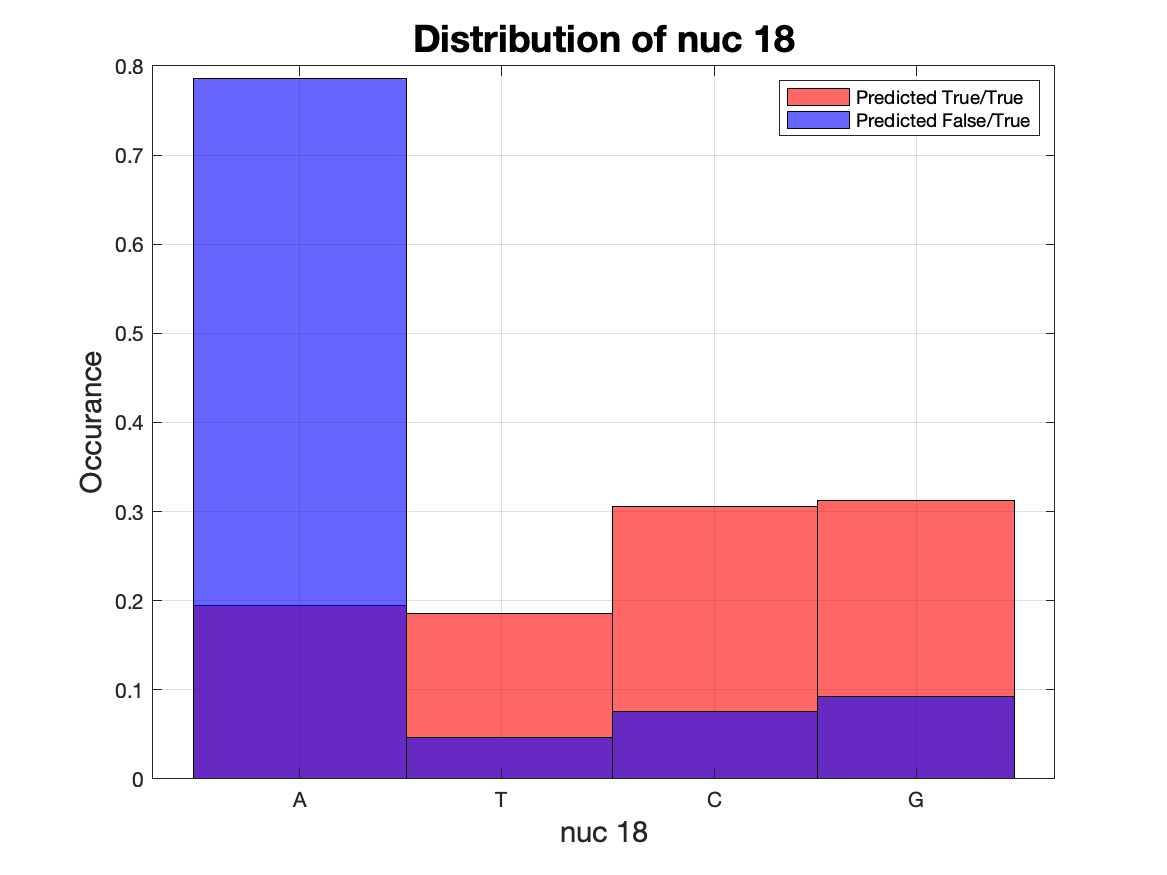}
		\caption{Distribution of the eighteenth nucleotide position.}\label{cnnfeatures18}
	\end{subfigure}

	\caption{Feature distribution for CNNs}\label{cnnfeatures}
\end{figure}

As seen in \autoref{cnnfeatures}, uniform and nonuniform distributions with respect to feature specific misclassifications are similar to that seen with SVM. For this reason, the nucleotide sensitivity across different positions in \autoref{cnnfeatures3} follows that observed in \autoref{svmfeatures8}. Conversely, \autoref{cnnfeatures18} displays a new trend observed in the true misclassifications from the eighteenth nucleotide feature onward to the last nucleotide feature: the distribution of true misclassifications does not follow the distribution of correctly classified sequences. In this case, the true correctly classified true entries in red display an almost uniform distribution, while the incorrectly classified true entries in blue display a distribution almost completely favored for A. This may indicate that at least toward the end of the ‘SEQ’ feature, the model failed to recognize a uniform distribution of nucleotides.

\section{Discussion}

When analyzing these models, one inquiry is whether a high FNR or FPR is less desirable. In the scope of this research, a false positive is defined as including a subsequence in the reference sequence when it should not have been included, and a false negative is defined as deciding not to include a subsequence in the reference sequence when it should have been included. In each optimal model, the FPR is much higher than the FNR. Initially, one would think that false positives are less desirable than false negatives. One of the arguments for this proposition is that fewer false positives allow the reference to be smaller in size. Since the size of the reference is not determined to be a major constraint, this need not be an issue.

Instead, the inclusion of these false positives increases the genetic variation in the reference if it is used recursively. Prior to restriction synthesis, a large reference would initially be generated from a random sequence. This reference would then be subsequently used in classifying more entries and so on and so forth. Paramount to the field of genomics, genetic variation in this context would allow for more subsequences to be analyzed and classified if this recursive process is applied with references containing a slightly higher FPR and slightly lower FNR.

The performance between these three methods pose another interesting discussion. The sensitivity of SVM, random forest, and CNN are 94.9\%, 92.7\%, and 91.4\% respectively. At least for this data set, SVM and random forest are far more consistent and accurate than CNN. Furthermore, the average specificity between all three vary greatly with random forest performing better than SVM and CNN. Of all three, SVM had the lowest specificity at 77.4\%, while CNN (82.4\%) trails behind random forest (85.7\%) by only 3.3\% and SVM behind random forest by 8.3\%. In then comparing SVM to random forest based on the sensitivity, the question becomes whether a 2.2\% increase in sensitivity is worth a 8.3\% decrease in specificity. Because of the previous discussion on the FPR and the FNR, the answer is definitively yes. To reiterate, a higher FPR coupled with a higher sensitivity does not impact the classification if the process is implemented recursively. For this reason, SVM is ranked ahead of random forest in these respects. Because both the specificity and sensitivity of CNN are below that of SVM and random forest, random forest is ranked ahead of CNN but below SVM. It is interesting, though, that all three techniques exhibited a lower specificity. One possible interpretation is that true cases are easier to differentiate.

SVM out performs both CNN and random forest in sensitivity. One hypothesis for why this might have been the case is due to the use of PCA to build representative components from the data. Surprisingly, all 16 unique features were able to be reduced to only two components, allowing for the classification to be observed in a two-dimensional space. Although SVM are the only technique to reach near the 95\% benchmark for sensitivity, the two other techniques did perform better than 90\% in that category. This indicates that these techniques may be improved, but it may require more work in the realm of feature selection, data preprocessing, and architecture design.

\footnotesize

\section*{Acknowledgment}
This work is supported by the National Science Foundation Award CCF-1937419 (RTML: Small: Design of System Software to Facilitate Real-Time Neuromorphic Computing).

\nocite{*}
%


\begin{thebibliography}{10}

\bibitem{ref1}
W.~{Tatum} and R.~E. {Hausman}.
\newblock {\em The Cell: A Molecular Approach, Sixth Edition}.
\newblock 2000.

\bibitem{reichard1988interactions}
Peter Reichard.
\newblock Interactions between deoxyribonucleotide and dna synthesis.
\newblock {\em Annual review of biochemistry}, 57(1):349--374, 1988.

\bibitem{ref2}
R.~A. {Hughes} and A.~D. {Ellington}.
\newblock {Synthetic DNA Synthesis and Assembly: Putting the Synthetic in
  Synthetic Biology}.
\newblock {\em {Cold Spring Harbor Perspectives in Biology}}, 9, January 2017.

\bibitem{caruthers1985gene}
Marvin~H Caruthers.
\newblock Gene synthesis machines: Dna chemistry and its uses.
\newblock {\em Science}, 230(4723):281--285, 1985.

\bibitem{ref3}
{National Human Genome Research Institute}.
\newblock {The Human Genome Project}, 2020.

\bibitem{ref4}
{Broad Institute}.
\newblock {CRISPR Timeline}, 2020.

\bibitem{ref5}
Nicholas Tang, Siying Ma, and Jingdong Tian.
\newblock {Chapter 1 - New Tools for Cost-Effective DNA Synthesis}.
\newblock In Huimin Zhao, editor, {\em Synthetic Biology}, pages 3 -- 21. 2013.

\bibitem{bigger1973recognition}
Cynthia~H Bigger, Kenneth Murray, and Noreen~E Murray.
\newblock Recognition sequence of a restriction enzyme.
\newblock {\em Nature New Biology}, 244(131):7--10, 1973.

\bibitem{ref7}
et~al. Anderson, JChristopher.
\newblock Bglbricks: A flexible standard for biological part assembly.
\newblock {\em Journal of Biological Engineering}, 4(1):1, 2010.

\bibitem{ref8}
{New England Biolabs}.
\newblock {Restriction Endonucleases}, 2020.

\bibitem{shetty2008engineering}
Reshma~P Shetty, Drew Endy, and Thomas~F Knight.
\newblock Engineering biobrick vectors from biobrick parts.
\newblock {\em Journal of biological engineering}, 2(1):5, 2008.

\bibitem{NCBI2020nuc}
{Gene [Internet]. Bethesda (MD): National Library of Medicine (US), National
  Center for Biotechnology Information}, 2020.

\bibitem{ref9}
{Nucleotide [Internet]}.
\newblock {National Library of Medicine (US), National Center for Biotechnology
  Information}, 2020.

\bibitem{ref23}
Johan~AK Suykens and Joos Vandewalle.
\newblock Least squares support vector machine classifiers.
\newblock {\em Neural processing letters}, 9(3):293--300, 1999.

\bibitem{ref24}
William~S Noble.
\newblock What is a support vector machine?
\newblock {\em Nature biotechnology}, 24(12):1565--1567, 2006.

\bibitem{ref25}
Baozhen Yao, Ping Hu, Mingheng Zhang, and Maoqing Jin.
\newblock A support vector machine with the tabu search algorithm for freeway
  incident detection.
\newblock {\em International Journal of Applied Mathematics and Computer
  Science}, 24(2):397--404, 2014.

\bibitem{ref11}
{J. Lever and M. Krzywinski and N. Altman}.
\newblock Principal component analysis.
\newblock {\em Nature Methods}, 14(7):641--642, 2017.

\bibitem{ref22}
Mahesh Pal.
\newblock Random forest classifier for remote sensing classification.
\newblock {\em International journal of remote sensing}, 26(1):217--222, 2005.

\bibitem{ref21}
Andy Liaw, Matthew Wiener, et~al.
\newblock Classification and regression by randomforest.
\newblock {\em R news}, 2(3):18--22, 2002.

\bibitem{ref12}
{L. Breiman and A. Cutler}.
\newblock {Random Forests - Classification Description}.
\newblock Berkeley University, 2020.

\bibitem{ref13}
Adele Cutler and John~R. Stevens.
\newblock Random forests for microarrays.
\newblock In {\em DNA Microarrays, Part B: Databases and Statistics}, volume
  411 of {\em Methods in Enzymology}, pages 422 -- 432. 2006.

\bibitem{ref15}
Ian Goodfellow, Yoshua Bengio, and Aaron Courville.
\newblock {\em Deep Learning}.
\newblock MIT Press, 2016.

\bibitem{ref20}
Yoshua Bengio, Ian Goodfellow, and Aaron Courville.
\newblock {\em Deep learning}, volume~1.
\newblock MIT press Massachusetts, USA:, 2017.

\bibitem{ref18}
LeCun, Yann and Bengio, Yoshua and Hinton, Geoffrey.
\newblock {\em Deep learning}, volume~521.
\newblock Nature, pages 436--444, 2015.

\bibitem{ref19}
J{\"u}rgen Schmidhuber.
\newblock Deep learning in neural networks: An overview.
\newblock {\em Neural Networks}, 61:85--117, 2015.

\end{thebibliography}

\end{document}